# Идентификация Квантовых Вихрей в Импульсном Пространстве

Н.В. Ларионов, В.М. Молчановский

Санкт-Петербургский государственный морской технический университет,

ул. Лоцманская, д. 3, Санкт-Петербург, 190121, Россия.

*e-mail:* larionov.nickolay@gmail.com

**Аннотация.** В работе теоретически исследуются квантовые вихри, образующиеся в результате надбарьерной ионизации двумерного атома водорода сверхкоротким лазерным импульсом. Анализируется поток вероятности в импульсном пространстве. При этом используется как стандартное выражение для потока, так и его «симметричный» аналог. Последний, вследствие чувствительности к фазе волновой функции, позволяет идентифицировать квантовые вихри.

**Ключевые слова:** квантовый вихрь, импульсное представление, поток вероятности

**Введение.** Одним из нетривиальных эффектов, возникающих при ионизации атома, является образование квантовых вихрей [1-10]. Эти вихри проявляют себя как возмущения в плотности вероятности для фотоэлектрона. Центр вихря представляет собой запрещенную для фотоэлектрона область, вокруг которой векторное поле потока вероятности имеет спиралевидную структуру.

В [4-6], используя как численный, так и аналитический подходы, мы изучали формирование и эволюцию квантовых вихрей, образующихся при надбарьерной ионизации двумерного атома водорода сверхкоротким лазерным импульсом. Аналитический подход был основан на решении уравнения Шредингера с помощью нестационарной теории возмущений. Была получена волновая функция фотоэлектрона, которая успешно использовалась для идентификации центров квантовых вихрей и дальнейшего сравнения с численными расчётами. Однако векторное поле потока вероятности с её помощью не анализировалось. Здесь мы восполним этот пробел, при этом будем использовать как стандартное выражение для потока в импульсном пространстве, так и альтернативное выражение, чувствительное к фазе волновой функции. Этот альтернативный «поток» был введен в работе [11] и является величиной «симметричной» по отношению к потоку, записанному в координатном представлении.

**Теоретическая модель.** Будем использовать систему атомных единиц, в которой уравнение Шредингера для двумерного атома водорода, взаимодействующего с лазерным импульсом, имеет вид

$$i\frac{\partial}{\partial t}\left|\Psi(t)\right\rangle = \left(\hat{H}_0 + \hat{V}\right)\left|\Psi(t)\right\rangle, \quad (1)$$

где $\hat{H}_0$ – гамильтониан свободного атома, $\hat{V} = -\hat{\mathbf{d}}\mathbf{F}(t)$ – оператор взаимодействия атома с электрическим полем лазера $\mathbf{F}(t)$, $\hat{\mathbf{d}} = -\hat{\mathbf{r}}$ – оператор дипольного момента.

Как и в предыдущих наших работах будем искать решение уравнения (1) в приближении сильного поля. Это означает, что в расчётах пренебрежём возбужденными состояниями атома, и не будем учитывать кулоновское воздействие на фотоэлектрон. Также будем считать, что изменения населенности основного (начального) состояния атома пренебрежимо малы. Тогда решение уравнения Шредингера (1) может быть записано в виде следующей суперпозиции

$$\left|\Psi(t)\right\rangle = \left|\Psi_{1,0}^{(0)}\right\rangle e^{-iE_1 t} + \sum_{m=0,\pm 1,\ldots} \int_0^\infty b_{k,m}(t)\left|\Psi_{k,m}^{(0)}\right\rangle e^{-iE_k t} k\, dk. \quad (2)$$

Здесь первое слагаемое соответствует основному состоянию атома с энергией $E_1 = -1/2$, характеризуемого вектором $\left|\Psi_{1,0}^{(0)}\right\rangle$. Нижние индексы «1,0» указывают на значения главного квантового числа $n = 1$ и проекции момента на ось $z$ $m = 0$. Второе слагаемое представлено суперпозицией векторов состояний фотоэлектрона $\left|\Psi_{k,m}^{(0)}\right\rangle$, которые будем описывать цилиндрическими волнами. Индексы «k,m» указывают на то, что это состояние характеризуется энергией $E_k = k^2/2 = \left(k_x^2 + k_y^2\right)/2$ и проекцией момента $m = 0, \pm 1, \pm 2,\ldots$

Так как нашей задачей является идентификация квантовых вихрей в импульсном пространстве, то перепишем (2) в соответствующем представлении

$$\Psi(\mathbf{k},t) \equiv \langle \mathbf{k} | \Psi(t) \rangle = \frac{2}{(k^2+1)^{3/2}} \Phi_0(\varphi_k) e^{-iE_1 t} + \sum_{m=0,\pm 1,\ldots} b_{k,m}(t)(-i)^{|m|} \Phi_m(\varphi_k) e^{-iE_k t}, \quad (3)$$

где $\mathbf{k} = (k, \varphi_k)$ - импульс электрона в полярной системе координат, $\Phi_m(\varphi_k) = e^{im\varphi_k}/\sqrt{2\pi}$ и были использованы явные выражения для волновых функций в импульсном представлении (см. [6])

$$\Psi_{1,0}^{(0)}(\mathbf{k}) \equiv \langle \mathbf{k} | \Psi_{1,0}^{(0)} \rangle = \frac{2\Phi_0(\varphi_k)}{(k^2+1)^{3/2}}, \quad \Psi_{k',m}^{(0)}(\mathbf{k}) \equiv \langle \mathbf{k} | \Psi_{k',m}^{(0)} \rangle = (-i)^{|m|} \frac{\delta(k'-k)}{k'} \Phi_m(\varphi_k). \quad (4)$$

Разложение (3) позволяет вывести замкнутое уравнение для неизвестных коэффициентов $b_{k,m}(t)$. Для этого подставим (3) (или (2)) в уравнение Шредингера (1), тогда, учитывая явный вид оператора взаимодействия в импульсном представлении $\hat{V} = \mathbf{F}(t) \cdot i\partial/\partial \mathbf{k} = F_x(t) i\partial/\partial k_x + F_y(t) i\partial/\partial k_y$, получаем

$$\frac{\partial b_{k,m}(t)}{\partial t} = \frac{(-i)}{2} \left( F_-(t) \delta_{m,1} + F_+(t) \delta_{m,-1} \right) \frac{6k \cdot e^{i\omega_{k1} t}}{(k^2+1)^{5/2}} + \\
+ \frac{(-i)^{|m-1|-|m|}}{2} F_-(t) \left( \frac{\partial}{\partial k} - ik \cdot t - \frac{m-1}{k} \right) b_{k,m-1}(t) + \frac{(-i)^{|m+1|-|m|}}{2} F_+(t) \left( \frac{\partial}{\partial k} - ik \cdot t + \frac{m+1}{k} \right) b_{k,m+1}(t), \quad (5)$$

где $\omega_{k1} = (k^2+1)/2$, $F_\pm(t) = F_x(t) \pm iF_y(t)$. Система (5) соответствует системе, полученной нами ранее в работе [6], однако здесь поляризация поля произвольна.

Решая систему (5) с помощью теории возмущений можно получить следующее выражение для части волновой функции (3), соответствующей непрерывному спектру

$$\tilde{\Psi}(\mathbf{k},t) = -i \left[ b_{k-1,10}^{(1)}(t) \Phi_{-1}(\varphi_k) + b_{k1,10}^{(1)}(t) \Phi_1(\varphi_k) \right] e^{-iE_k t} + \\
+ b_{k0,10}^{(2)}(t) \Phi_0(\varphi_k) e^{-iE_k t} - \left[ b_{k-2,10}^{(2)}(t) \Phi_{-2}(\varphi_k) + b_{k2,10}^{(2)}(t) \Phi_2(\varphi_k) \right] e^{-iE_k t}, \quad (6)$$

где верхний индекс у амплитуд соответствует порядку теории возмущений (см. [5,6]), а добавленный нижний индекс «10» указывает на начальное связанное состояние электрона.

Знак тильда над $\Psi$ подчеркивает (в [5,6] было принято обозначение $\tilde{\Psi}(\mathbf{k},t) = b(\mathbf{k},t)e^{-iE_k t}$), что связанное состояние опущено, т.е. в нашем дальнейшем рассмотрении интерференцией между начальным и конечными состояниями электрона пренебрегается.

Нашей задачей является идентификация квантовых вихрей в импульсном пространстве. Для этого будем использовать следующие два определения для потока вероятности в импульсном пространстве $\mathbf{k}$

$$\mathbf{j}(\mathbf{k},t) = \mathbf{k}\left|\tilde{\Psi}(\mathbf{k},t)\right|^2, \quad \overline{\mathbf{j}}(\mathbf{k},t) = -\frac{1}{2i}\left[\tilde{\Psi}^*(\mathbf{k},t)\nabla_k \tilde{\Psi}(\mathbf{k},t) - \tilde{\Psi}(\mathbf{k},t)\nabla_k \tilde{\Psi}^*(\mathbf{k},t)\right]. \quad (7)$$

Здесь $\mathbf{j}(\mathbf{k},t)$ - стандартное выражение для потока вероятности в импульсном пространстве, а $\overline{\mathbf{j}}(\mathbf{k},t)$ - «симметричный» поток вероятности [11], где $\nabla_k = \partial/\partial \mathbf{k}$.

Как показано нами ранее [5,6], квантовые вихри, появляющиеся в процессе ионизации атома, обусловлены интерференцией состояний фотоэлектрона. Следовательно, для того чтобы корректно описать эти вихри необходимо знание о фазе волновой функции. Выделим эту фазу в найденной волновой функции (6): $\tilde{\Psi}(\mathbf{k},t) = \left|\tilde{\Psi}(\mathbf{k},t)\right|e^{i\chi(\mathbf{k},t)}$, где $\chi(\mathbf{k},t)$ - фаза. Если подставить $\tilde{\Psi}(\mathbf{k},t)$ в стандартное определение потока вероятности $\mathbf{j}(\mathbf{k},t)$, то фаза $\chi(\mathbf{k},t)$ исчезнет, а идентификация квантового вихря будет возможна, в основном, по нулям волновой функции. Другое дело для альтернативного потока $\overline{\mathbf{j}}(\mathbf{k},t)$. В самом деле, подставив $\tilde{\Psi}(\mathbf{k},t)$ в выражение для $\overline{\mathbf{j}}(\mathbf{k},t)$ получаем $\overline{\mathbf{j}}(\mathbf{k},t) = -\left|\tilde{\Psi}(\mathbf{k},t)\right|^2 \nabla_k \chi(\mathbf{k},t)$.

**Результаты расчётов и обсуждения.** Напряженность поля лазерного импульса промоделируем следующим выражением

$$\mathbf{F}(t) = \mathbf{e}_x F_0 \cos(\omega t)\left[\theta(T-t) - \theta(-t)\right], \quad (8)$$

где $\theta(x)$ - функция Хэвисайда, $F_0$ - постоянная амплитуда, $\omega$ - частота, $T$ – длительность импульса, $\mathbf{e}_x$ - единичный вектор в направлении оси $x$.

Значения параметров импульса выберем близкими к тем, для которых ранее были идентифицированы квантовые вихри [4-6]: $F_0 = 0.6$, $\omega = \pi$, $T = 2$. Будем интересоваться установившимся решением, т.е. исследовать $\tilde{\Psi}(\mathbf{k},t)$ (6) на временах $t > T$.

Сделаем несколько замечаний относительно выбранных параметров. Значение амплитуды поля берется таким, чтобы преобладала надбарьерная ионизации, но при этом $F_0 < 1$. Это позволяет оставаться в пределах используемых теоретических приближений. На «масштаб» вихря сильно влияет продолжительность импульса $T$ [4]. Как показывают расчёты при $T < 1$ вихрь не успевает сформироваться и его невозможно идентифицировать. При переходе к большим значениям $T > 10$ запрещённая область постепенно «размазывается» и в конечном итоге вихрь теряет свою индивидуальность, а «картина» становиться похожа на ту, которая имеет место при ионизации монохроматическим полем.

На рисунке 1 представлены графики распределения по импульсам фотоэлектрона $\left|\tilde{\Psi}(k_x,k_y,t>T)\right|^2$ 1a,b) (для более четкого отображения график строится для $\ln\left|\tilde{\Psi}\right|^2$) и соответствующие векторные поля для стандартного потока вероятности $\mathbf{j}(k_x,k_y,t>T)$ 1c) и «симметричного» потока $\overline{\mathbf{j}}(k_x,k_y,t>T)$ 1d).

Для выбранных параметров наблюдаются два симметричных вихря с центрами в точках $k_x \approx 0, k_y \approx \pm 2.3$. На Рис.1a) эти центры выделены кружочками, а на Рис.1b) центр вихря с положительной координатой $k_y \approx 2.3$ показан в увеличенном масштабе.

Рисунок 1c) иллюстрирует векторное поле соответствующее стандартному определению потока вероятности в импульсном пространстве. Можно видеть, что в окрестности центра вихря $k_x \approx 0, k_y \approx 2.3$ векторное поле никак не выделено и совпадает с усреднённым полем скоростей в данной области импульсного пространства. Только более тёмный фон на графике указывает на наличие запрещённой области для фотоэлектрона.

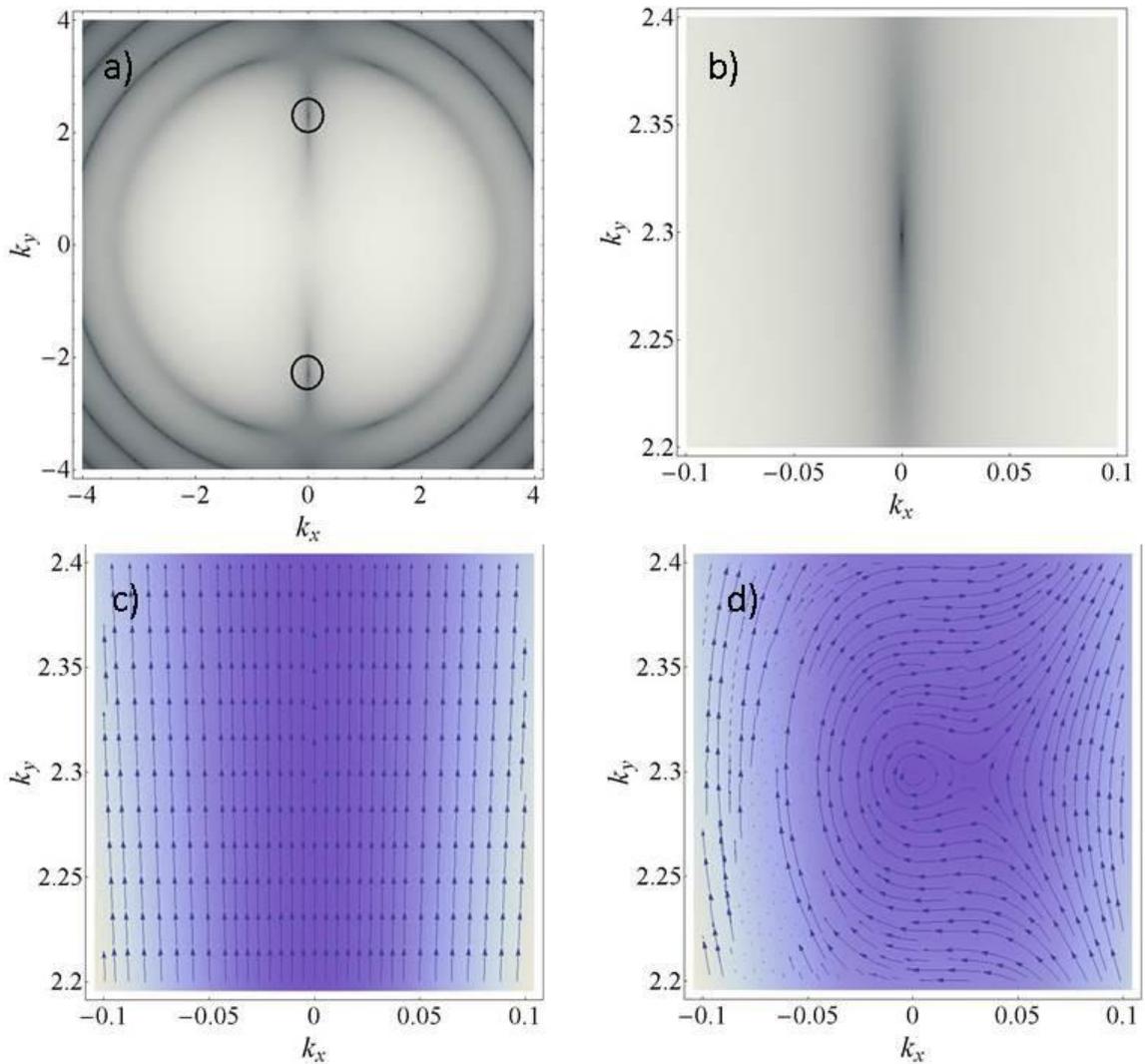

Рис. 1. a) и b): распределение по импульсам фотоэлектрона $\ln\left(\left|\tilde{\Psi}\left(k_x,k_y,t>T\right)\right|^2\right)$. Векторные поля для стандартного потока вероятности $\mathbf{j}\left(k_x,k_y,t>T\right)$ c) и «симметричного» потока $\overline{\mathbf{j}}\left(k_x,k_y,t>T\right)$ d).

Совсем по-другому выглядит векторное поле, полученное с помощью «симметричного» потока Рис.1d). Отчётливо видно, что вокруг нуля волновой функции поле $\overline{\mathbf{j}}\left(k_x,k_y,t>T\right)$ имеет вихревую структуру, близкую к соленоидальной.

**Заключение.** Таким образом, в данной статье показано, что с помощью «симметричного» выражения для потока вероятности $\overline{\mathbf{j}}(\mathbf{k},t)$ (7), введённого в работе [11], возможно идентифицировать квантовые вихри в импульсном пространстве.